\documentclass[twocolumn,preprintnumbers,elsart]{revtex4}
\usepackage{amsmath}
\usepackage{booktabs}
\usepackage{cases}
\usepackage{mathrsfs}
\usepackage{graphicx}
\usepackage{dcolumn}
\usepackage{bm}
\usepackage[center]{subfigure}
\usepackage{color}
\usepackage{float}

\begin{document}
\title{Photonic Spin-Orbit Coupling Induced by Deep-Subwavelength Structured Light}
\author{Xin Zhang$^{1,2}$, Guohua Liu$^{1,2}$, Yanwen Hu$^{1,2,3*}$, Haolin Lin$^{1,2}$, Zepei Zeng$^{1,2}$, Xiliang Zhang$^{1,2}$, Zhen Li$^{1,2,3}$, Zhenqiang Chen$^{1,2,3}$, and Shenhe Fu$^{1,2,3}$}
\email{huyanwen@jnu.edu.cn;fushenhe@jnu.edu.cn}
\address{
$^{1}$Department of Optoelectronic Engineering, Jinan University, Guangzhou 510632, China \\
$^{2}$Guangdong Provincial Key Laboratory of Optical Fiber Sensing and Communications, Guangzhou 510632, China \\
$^{3}$Guangdong Provincial Engineering Research Center of Crystal and Laser Technology, Guangzhou 510632, China}

\begin{abstract}
\noindent We demonstrate both theoretically and experimentally beam-dependent photonic spin-orbit coupling in a two-wave mixing process described by an equivalent of the Pauli equation in quantum mechanics. The considered structured light in the system is comprising a superposition of two orthogonal spin-orbit-coupled states defined as spin up and spin down equivalents. The spin-orbit coupling is manifested by prominent pseudo spin precession as well as spin-transport-induced orbital angular momentum generation in a photonic crystal film of wavelength thickness. The coupling effect is significantly enhanced by using a deep-subwavelength carrier envelope, different from previous studies which depend on materials. The beam-dependent coupling effect can find intriguing applications; for instance, it is used in precisely measuring variation of light with spatial resolution up to 15 nm.  
\end{abstract}
\maketitle
\section{Introduction}
\noindent Spin-orbit coupling (SOC), which refers to interaction of a quantum particle’s spin with its momentum, is a fundamentally important concept. It has been extensively investigated in condensed matter physics \cite{Winkler2003,Goldman2018}, atomic and molecular physics \cite{Spielman2011,Spielman2013} and contributes to exciting phenomena such as the spin Hall effect \cite{Kato2004} and topological insulators \cite{Hasan2010,Taylor2011}. Analogous photonic SOC is also demonstrated in a variety of settings \cite{Zayats2015}. The photonic SOC refers to an interaction between the momentum of light, which also includes spin angular momentum and orbital angular momentum (SAM and OAM). Whereas the SAM is associated with photon circular polarization \cite{Padgett2002}, the OAM is relevant to a helical wavefront of light characterized by a topological number $\ell$ \cite{Allen1992}. The photonic SOC is crucial for the optical Hall effects \cite{Nagaosa2004,Glazov2005,Kwiat2008,Fu2019}, spin-to-orbital angular momentum conversions \cite{Marrucci2006,Chiu2007}, spin-orbit photonic devices \cite{Gorodetski2008,Shitrit2013,Wang2021}, etc. \\
\indent The SOC can be engineered in appropriately designed materials. For examples, engineering a tensional strain in graphene shifts the electronic dispersions and induces a controllable vector potential for the electronic SOC \cite{Crommie2010,Low2010,Juan2011,Guinea2010,Bolotin2009}. Analogous strategy can be applied to engineer the photonic SOC, by using strained evanescently coupled waveguide arrays \cite{Rechtsman2013,Lumer2019}. Other approaches for manipulating the photonic SOCs are demonstrated by appropriately designing microcavities \cite{RO2011,Longhi2013,Fang2012,Yu2012,SF2013,Szczytko2021}, metamaterials \cite{Liu2015,Sala2015,Zhang2023,Lu2023}, photonic crystals \cite{Yesharim2022,Karnieli2018,Liu2023}, twisted optical fibers \cite{Chiao1986,Liberman1992}, dual-core waveguides \cite{Torner2015,Malomed2016}, etc. The resultant SOCs are material-dependent, determined by geometric configurations of the materials which are often difficult to be tuned once fixed by designs. As a consequence, a tunable photonic SOC process remains elusive. Recently, several engineered photonic SOC schemes have been reported, by either embedding a strained honeycomb metasurface inside a cavity waveguide \cite{Mann2020} or using an optical cavity filled with controllable liquid crystals \cite{Szczytko2019}. However, the resultant photonic SOCs remain material-dependent. \\
\indent In this work, we report theoretically and experimentally a new mechanism for engineering the photonic SOC. We demonstrate this by exploiting analogy between quantum description of a spin-1/2 system and a spin-orbit Hamiltonian derived for structured light in a photonic crystal. The obtained Hamiltonian is closely relevant to structured light, which means that the SOC can be engineered by controlling carrier envelope rather than the structures of materials. Strong SOC is achieved by using deep-subwavelength structured light, as manifested by clear pseudo spin precessions. Although the structured light has been extensively investigated in recent year \cite{Forbes2021,Shen2019,Fu2023,Femius2015,Hu2021,Fu2020}, the dependence of the photonic SOC on its spatial structure remains unnoticed.
\section{Theoretical model}
\begin{figure*}[t]
	\centering
	\includegraphics[width=16cm]{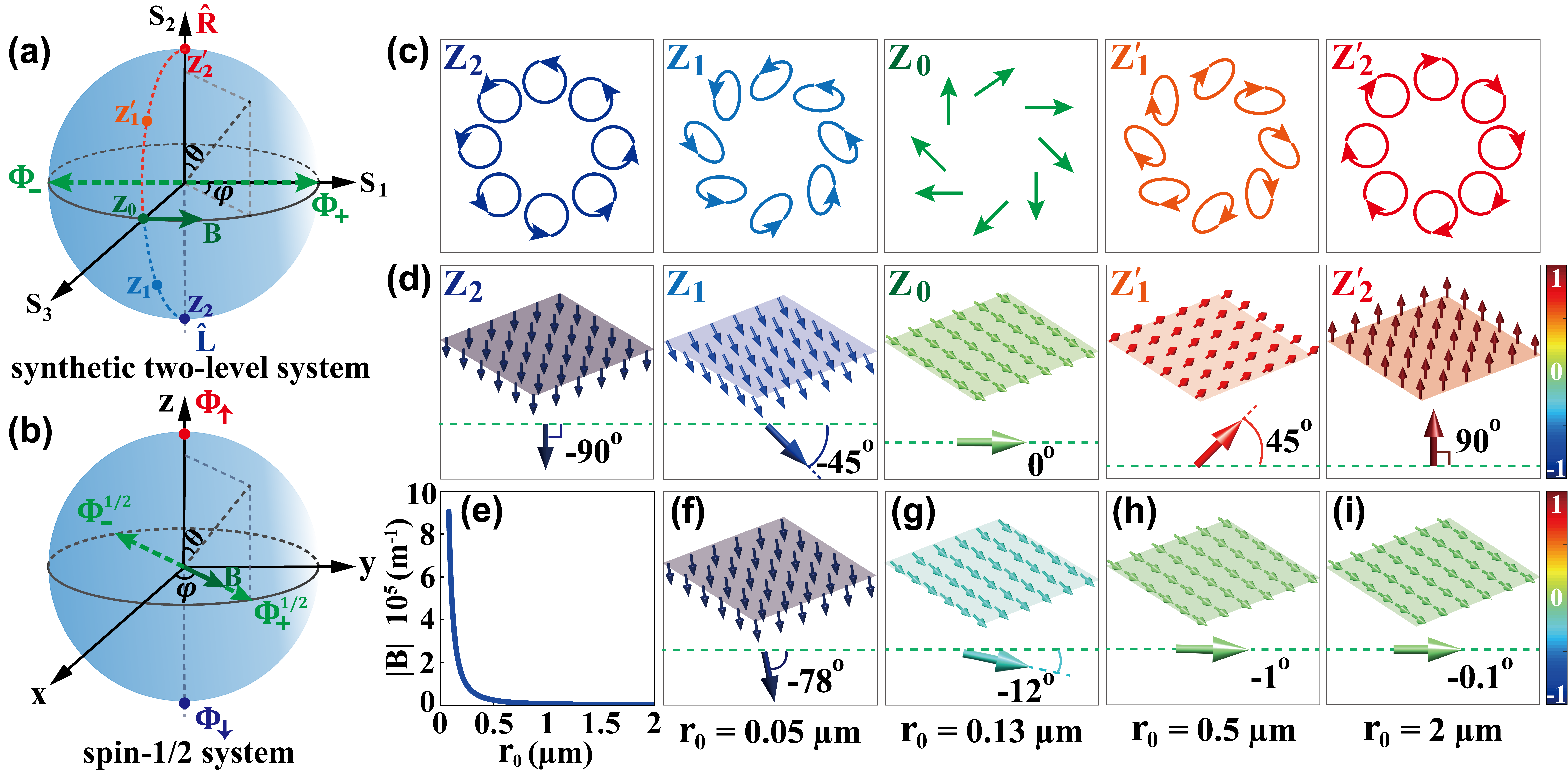}
	\caption{(a) Geometrical representation of spin precession in the presence of SOC. $\hat{R}$ and $\hat{L}$ define the spin-up and spin-down equivalents in the $B_2$ direction, respectively; whereas $\Phi_+$ and $\Phi_-$ denotes two spin eigenstates in the direction of synthetic field \textbf{B}. Spin precession is initiated by a mixing spin $\Phi=1/\sqrt{2}(\Phi_++i\Phi_-)$ located at $z_0$. (b) Bloch-sphere representation of spin-1/2 system, in the presence of external field \textbf{B}. $\Phi_\uparrow$ and $\Phi_\downarrow$ are spin up and spin down in the $z$ direction, while $\Phi_+^{1/2}$ and $\Phi_-^{1/2}$ denotes eigenstates of the system, corresponding to direction of \textbf{B}. (c) Polarization states mapped on a longitude line of the first-order ($\ell=1$) sphere in (a). (d) Corresponding spin vectors to (c). (e) SOC strength as a function of beam width $r_0$. (f)-(i) Theoretical results for the spin vectors under actions of LG beam with different widths.}
\end{figure*}
\indent We consider a two-wave mixing process involving two interacting photonic states. The SOC takes place in a crystal, represented by its principal refractive index: $n_x$, $n_y$, and $n_z$. With an approximation of the slowly varying envelope along optical axis $z$, a coupled-wave equation for the process is given by \cite{Guo2018}
\begin{eqnarray}
\begin{aligned}
2i\beta_{x}\frac{\partial E_{x}}{\partial z}+\frac{n_{x}^2}{n_{z}^2}\frac{\partial^2E_{x}}{\partial x^2}+\frac{\partial^2E_{x}}{\partial y^2}=\gamma_{y}\frac{\partial^2E_{y}}{\partial y\partial x}\exp\left(+i\Delta\beta\cdot z\right) \\
2i\beta_{y}\frac{\partial E_{y}}{\partial z}+\frac{n_{y}^2}{n_{z}^2}\frac{\partial^2E_{y}}{\partial y^2}+\frac{\partial^2E_{y}}{\partial x^2}=\gamma_{x}\frac{\partial^2E_{x}}{\partial x\partial y}\exp\left(-i\Delta\beta\cdot z\right)
\end{aligned}
\end{eqnarray}
where $E_{x,y}$ are linearly polarized fields, and $\beta_{x,y}=k_{0}n_{x,y}$ denote their propagation constants. $k_{0}=2\pi/\lambda$ is free-space wavenumber with $\lambda$ being the wavelength. $\Delta\beta=\beta_{y}-\beta_{x}$ is a phase mismatch. We define $\gamma_{x,y}=1-n_{x,y}^2/n_{z}^2$ as coupling parameters, related to crystal's polarity. The derivatives $\nabla^2_{xy}$ and $\nabla^2_{yx}$ in Eq. (1) stem from the non-zero term $\nabla\cdot\textbf{E}\neq0$, featuring origin of the SOC \cite{Liberman1992}. \\
\indent To address the rapid oscillation terms $\exp(\pm i\Delta\beta\cdot z)$, we transform the wave equation to a rotating form, by defining
\begin{eqnarray}
	\begin{aligned}
		 E_x=\tilde{A}_x\exp(+i{\Delta}\beta\cdot z/2) \\
		 E_y=\tilde{A}_y\exp(-i{\Delta}\beta\cdot z/2)
	\end{aligned} 
\end{eqnarray}
respectively. Thus a Hamiltonian of the system is written as
\begin{equation}
\textbf{H}=\frac{1}{2\bar{\beta}}\left[
\begin{array}{cc}
-\nabla_\perp^2+\bar{\gamma}\nabla^2_{xx}, 0 \\
0, -\nabla_\perp^2+\bar{\gamma}\nabla^2_{yy}
\end{array}
\right]+\left[
\begin{array}{cc}
\Delta\beta/2, \bar{\gamma}\nabla^2_{yx}/(2\bar{\beta}) \\
\bar{\gamma}\nabla^2_{xy}/(2\bar{\beta}), -\Delta\beta/2
\end{array}
\right]
\end{equation}
where $\nabla_\perp^2=\nabla^2_{xx}+\nabla^2_{yy}$ denotes the Laplace operator. We have assumed shallow crystal birefringence, namely $\bar{\beta}\approx(\beta_x+\beta_y)/2$ and $\bar{\gamma}\approx(\gamma_x+\gamma_y)/2$. The second term in Eq. (3), which includes the derivative operators, couples the two polarization components. It means that the SOC is related to spatial structure of light.  \\
\indent We study the beam-dependent SOC in a synthetic two-level spin-orbit system. We define right and left circularly polarized vortex states as spin up and spin down equivalents in the $z$ direction. They are written as \cite{HuOL,Naidoo2016,Milione2011}:
\begin{eqnarray}
	\begin{aligned}
		\hat{R}=\exp(+i\ell\phi)(\hat{x}-i\hat{y})/\sqrt{2} \\
		\hat{L}=\exp(-i\ell\phi)(\hat{x}+i\hat{y})/\sqrt{2}
	\end{aligned} 
\end{eqnarray}
respectively, where $\hat{x}$ and $\hat{y}$ are unit vectors and $\phi=\arctan(y/x)$. Since the pseudo spins are defined in the circular basis, the Hamiltonian is modified by a transformation from the cartesian coordinate to the circular basis, yielding 
\begin{equation}
\textbf{H}'=\frac{\bar{\gamma}-2}{4\bar{\beta}}\left[\begin{array}{cc}\nabla_\perp^2, 0 \\
		0, \nabla_\perp^2
	\end{array}
	\right]+\left[
	\begin{array}{cc}
		0, \Delta\beta/2-i\bar{\gamma}\nabla^2_{yx}/(2\bar{\beta}) \\
		\Delta\beta/2+i\bar{\gamma}\nabla^2_{yx}/(2\bar{\beta}), 0
	\end{array}\right]
\end{equation}
Given an overall field $\tilde{\textbf{A}}=\tilde{A}(x,y,z)(\Phi_R\hat{R}+\Phi_L\hat{L})$, where $\Phi_R$ and $\Phi_L$ are weights on $\hat{R}$ and $\hat{L}$, respectively, we reduce Eq. (1) to the Schr$\ddot{\text{o}}$dinger-like (Pauli) form
\begin{equation}
	i\frac{\partial\Phi(z)}{\partial z}=\left(\frac{1}{2M}
	\textbf{P}_{\perp}^2{\tilde{A}}-\frac{1}{2}\sigma\cdot\textbf{B}\right)\Phi(z)
\end{equation}
where $\Phi=(\Phi_R,\Phi_L)^{\text{T}}$, $\textbf{P}_{\perp}^2=[-\nabla_\perp^2, 0; 0, -\nabla_\perp^2]$, and $M=2\bar{\beta}\tilde{A}/(2-\bar{\gamma})$. Here $\sigma$ is the Pauli matrix vector. The SOC is described by a term $-\sigma\cdot$\textbf{B}, where $B_1=-\bar{\gamma}\nabla^2_{xy}\tilde{A}/(\bar{\beta}\tilde{A})$, $B_2=0$, and $B_3=-\Delta\beta$. It is analogous to a coupling form which describes interaction between a particle's spin and its angular momentum in a moving frame \cite{Winkler2003,Goldman2018}. More details refer to Appendix A. Since $B_2$ is zero, the vector \textbf{B} lies on the purely transverse $B_1B_3$ plane. The SOC Hamiltonian admits eigenstates that point to the vector \textbf{B} and comprise an equal superposition of $\hat{R}$ and $\hat{L}$, written as
\begin{eqnarray}
\begin{aligned}
\Phi_+=1/\sqrt{2}\left[\hat{R}+\exp(i\varphi)\hat{L}\right] \\
\Phi_-=1/\sqrt{2}\left[\hat{R}-\exp(i\varphi)\hat{L}\right]
\end{aligned} 
\end{eqnarray}
Figure 1(a) visualize the eigenstates and the pure states ($\hat{R}$ and $\hat{L}$) in the Poincar$\acute{\text{e}}$ sphere, showing close analogies to Bloch-sphere representation of the spin-1/2 system \cite{Scully1987,Berry1987} [Fig. 1(b)]. The spin states exhibit cylindrically symmetric polarization distributions. As illustration, Fig. 1(c) displays typical polarizations of states mapped onto a longitude line in the first-order ($\ell=1$) sphere; while Fig. 1(d) depicts their corresponding spin vectors, represented by an angle $\arccos (S_2)$, where $S_2$ is value of polarization ellipticity. Since the state exhibits identical polarization ellipticity in the transverse plane, the resultant spin vectors are homogeneous.\\ 
\indent The SOC term shows a dynamical effect, caused by the propagation-variant envelope $\tilde{A}$. This shows sharp contrast to conventional ones which are usually being independent terms. However, if optical diffraction is neglected, the dynamical behavior disappears and the SOC strongly relies on the envelope. In this scenario, a relevant beam parameter becomes an important degree of freedom for engineering the SOC. This is demonstrated in Fig. 1(e), showing close relationship between SOC strength and beam width in the phase-matching condition ($\Delta\beta$=0). Here the Laguerre-Gaussian (LG) envelope is considered as: 
\begin{equation}
	\tilde{A}(r)=\frac{r}{r_0}\exp(-\frac{r^2}{r_0^2})
\end{equation}
where $r=(x^2+y^2)^{1/2}$, and $r_0$ features the beam width. At the deep-subwavelenth region ($r_0<\lambda/2$), the SOC strength is rapidly increasing with a slight decrease of $r_0$. It becomes relatively negligible when $r_0>\lambda$. This relation suggests that shrinking light to deep-subwavelength scale significantly enhances the SOC. Although the derivations are based on the slowly varying envelope approximation, the model can be applied to deep-subwavelength regime at the early stage of spin evolution.\\
\indent To demonstrate the deep-subwavelength-induced SOC, we set the coupling length to be only one cycle ($z=\lambda$), such that a moderate SOC cannot cause obvious spin transport phenomenon. On the other hand, the SOC strength can be maintained during beam propagation, due to the short coupling length. This results in an adiabatic spin evolution, represented as a spin precession around \textbf{B}, i.e.,
\begin{equation}
	\frac{d\mathbf{S}}{dz}=\mathbf{B}\times\mathbf{S}
\end{equation}
where $\mathbf{S}=(S_1, S_2, S_3)$ is the state vector defined as $S_h=\Phi^{\dag}\sigma_h\Phi$ ($h=1, 2, 3$). The spin vector is therefore described by $S_2$. We initiate the spin precession from a mixing state: $\Phi=1/\sqrt{2}[\Phi_++i\Phi_-]$. Figure 1(f)-1(i) display theoretical distributions of the spin vectors for different beam parameters. Evidently, for $r_0$=0.05 $\mu$m, the spin rotates to an angle about -78$^{\text{o}}$; By comparison, increasing the parameter to $r_0=0.13$ $\mu$m causes less significant spin precession, manifested by a spin rotation angle about -12$^{\text{o}}$. This indicates that the SOC strongly depends on the carrier envelope. Figure 1(h, i) show that a moderate SOC induced by the relatively larger envelope cannot cause spin precession.
\section{Experimental results and discussion}
\begin{figure}[t]
	\centering
	\includegraphics[width=8.6cm]{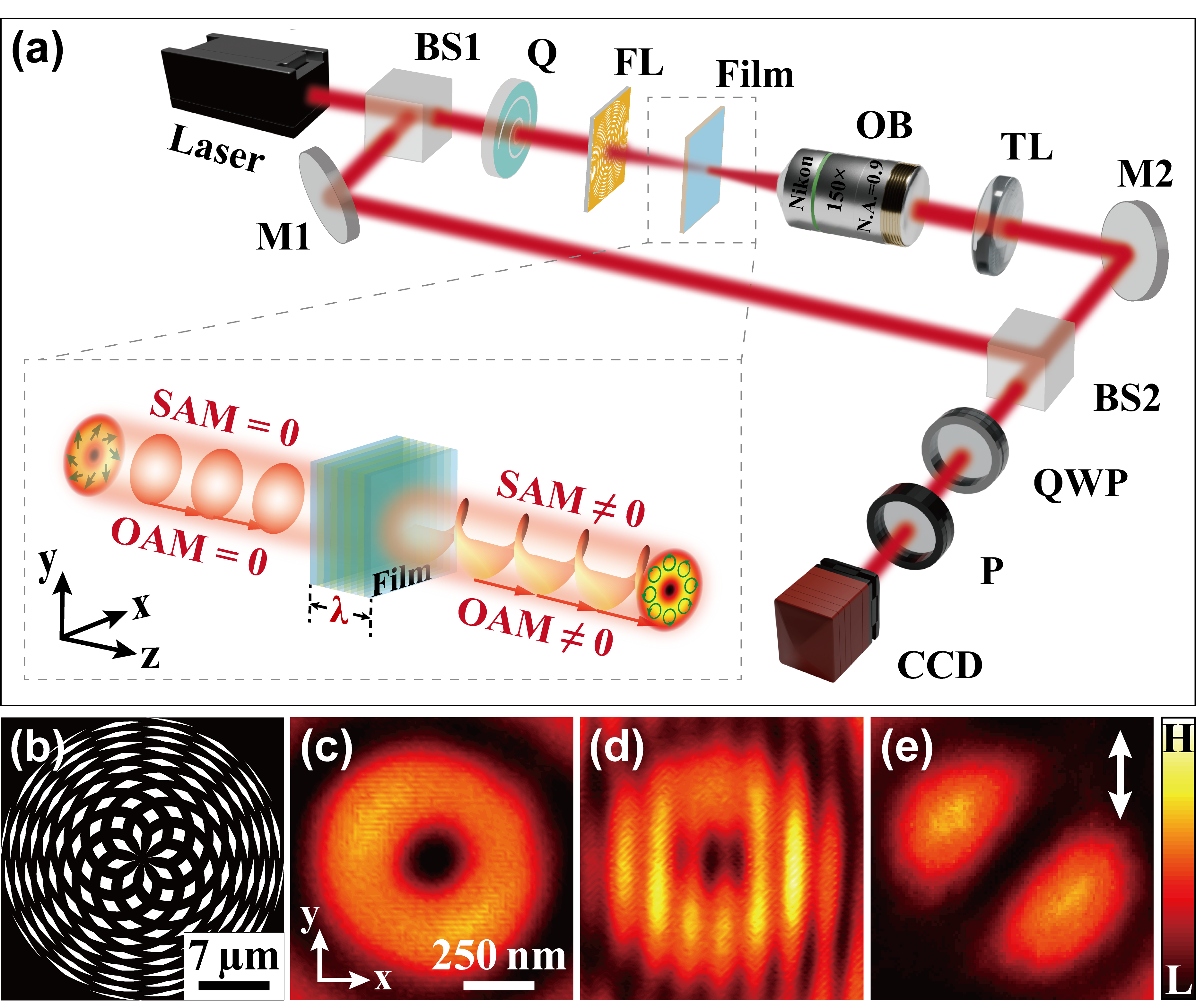}
	\caption{(a) Experimental setup. BS: beam splitter; Q: q-plate; FL: flat lens; M: mirror; OB: objective lens; TL: tube lens; QWP: quarter wave plate; P: polarizer; CCD: charge coupled device. The laser is operating at wavelength of $\lambda=632.8$ nm. The insert in (a) shows that an equatorial mixing spin with equal weight on $\Phi_R$ and $\Phi_L$ is adiabatically converted to a pure spin down in the presence of the SOC. (b) Layout of the 60-nm-thick flat lens with NA=0.87. (c) Intensity distribution of the LG beam at the focal plane ($z_f$) of the flat lens. (d) Plane-wave interference and (e) $y$-polarization component of beam at $z_f$, indicating a generation of the expected spin state $\Phi=1/\sqrt{2}(\Phi_++i\Phi_-)$. The scale bar in (c-e) is 250 nm. In color bar, L: low; H: high.}
\end{figure}
\indent Experiments are carried out to confirm the predictions. A crucial ingredient is to generate the required spin-orbit state at the deep-subwavelength scale. This is challenging since the incident state cannot maintain its property after tightly focused by the high-numerical-aperture (NA) objective lens \cite{Dorn2003,Wang2008,Xie2014}. To overcome this problem, we fabricate a topology-preserving high-NA flat lens (the thickness is 60 nm) according to a technique reported in \cite{Zhang2022}. The flat lens [the layout is shown in Fig. 2(b)] has a NA up to 0.87 and a focal length of $z_f=8$ $\mu$m. A system comprising an objective lens (150$\times$, NA=0.9) and a tube lens is utilized to characterize the flat lens, see Fig. 2(a). Figure 2(c) presents recorded intensity distribution of light at the focal plane. The focused LG beam exhibits a parameter of $r_0\simeq 0.32$ $\mu$m. The recorded regular interference [(Fig. 2(d)] and $y$-polarization component [(Fig. 2(e)] suggest that the expected initial spin is generated. Theoretical derivation about topology-preserving property of the flat lens (Appendix B) further confirms the generation. \\
\indent An experimental setup is built for measuring the spin procession. A linearly polarized He-Ne laser ($\lambda=632.8$ nm) is divided by a beam splitter. A q-plate with a charge of $q=1/2$ is applied to transform the beam into expected spin state carrier by the LG envelope. The purity of the spin state from the q-plate is measured as 95.2\% (Appendix C). The LG beam is focused into deep-subwavelength region by the flat lens. A c-cut lithium niobate crystal film ($\bar{\gamma}=-0.08$) with a thickness about one wavelength is placed at the focal plane. The emerging beam, in the presence of the SOC, is expected to accumulate a non-trivial spin phenomenon [see the insert in Fig. 2(a)]. 
\begin{figure}[t]
	\centering
	\includegraphics[width=8.6cm]{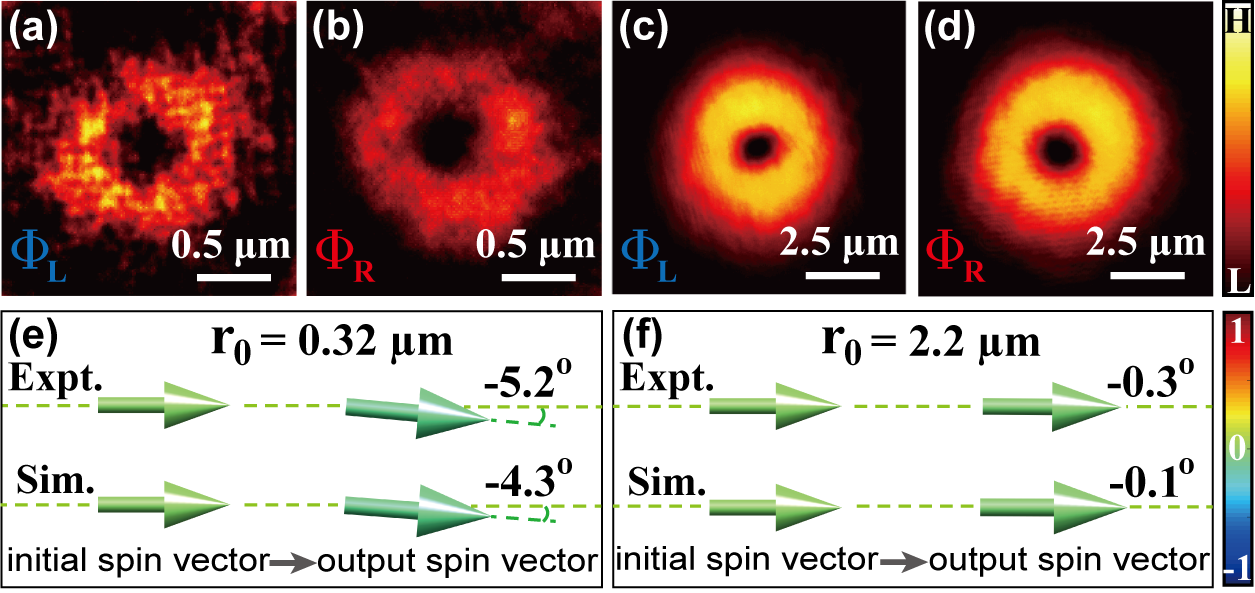}
	\caption{Experimental observation of spin rotation induced by the deep-subwavelength LG beam ($r_0=0.32$ $\mu$m), as manifested by spin angular momentum conversion (flipping) from right-handed one (b) to the left-handed one (a). In comparison, a larger LG beam parameter $r_0$=2.2 $\mu$m is considered, resulting in balanced left-handed (c) and right-handed (d) components. (e, f) The measured spin vectors before and after the crystal film, for (e) $r_0=0.32$ $\mu$m, and (f) $r_0=2.2$ $\mu$m. In the color bar (d), L: low; H: high.}
\end{figure} \\
\indent Figure 3 presents measurements confirming the spin precession. Since the spin is relevant to the circular polarization, we measure the right- (spin $\uparrow$) and left-handed (spin $\downarrow$) circular polarization components. These are achieved by rotating a quarter wave plate to an angle of $-\pi/4$ and $+\pi/4$ with respect to $x$ axis, respectively, while inserting a linear polarizer in front of the camera. Figure 3(a) and 3(b) depicts intensity distributions of $\Phi_L$ and $\Phi_R$, respectively. The measured $\Phi_L$ component is stronger than the $\Phi_R$ one, indicating a spin precession toward south pole of the sphere. Figure 3(e) shows the measured spin rotation by an angle of -5.2$^{\text{o}}$, compared to the initial one \cite{SR}. This approximately matches to the simulated result. However, for a larger parameter ($r_0=2.2$ $\mu$m), the $\Phi_L$ and $\Phi_R$ components are approximately identical [Fig. 3(c, d)], meaning that the induced SOC is insufficient to flip the spin [Fig. 3(f)]. Slight difference between the experiment and theory can be mainly attributed to the imperfect LG envelope that is closely relevant to the derivative operator $\nabla_{xy}^2$ \cite{SR}. \\
\indent We observe non-trivial spin-precession phenomenon, manifested by a generation of the photonic OAM. Initially, both the SAM and OAM of state at the equator are zero. Under the action of the SOC, its intrinsic OAM and SAM are separated simultaneously. This non-trivial phenomenon is observed in Fig. 4(a), showing a clear dislocation in the plane-wave interference fringes for the deep-subwavelength LG beam. This is a manifestation of wavefront helicity with a topological charge being $\ell=1$. The spin precession accompanied by the OAM generation confirms the phenomenon of spin-orbit separation. This effect becomes negligible for larger envelope, since the spin remains at its original position, as indicated by the regular interference fringes [Fig. 4(c)]. Theoretical results correspondingly shown in Fig. 4(b) and 4(d) are in accordance with the measurements.\\
\begin{figure}[t]
	\centering
	\includegraphics[width=8.6cm]{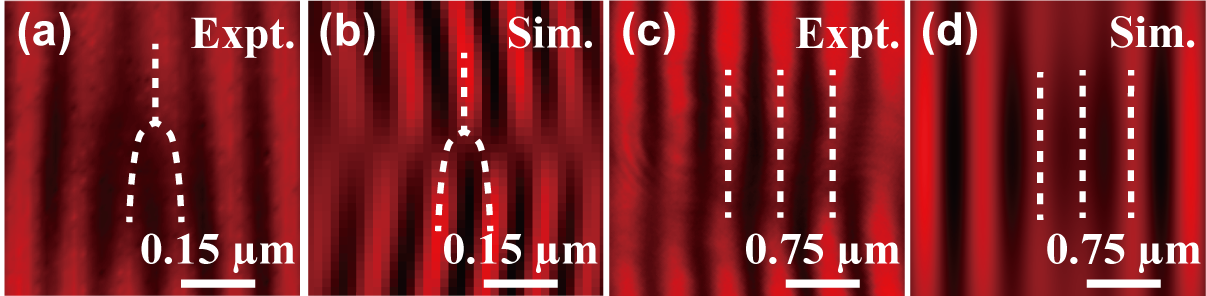}
	\caption{Observation of the orbital-angular-momentum state induced by spin precession. (a, c) The experimentally measured plane-wave interference patterns, for two different LG beam parameters: (a) $r_0=0.32$ $\mu$m, and (c) $r_0=2.2$ $\mu$m. (b, d) The simulated [based on Eq. (1)] interference patterns corresponding to the measurements in (a, c). Experimental conditions are kept the same as those in Fig. 3. }
\end{figure}
\begin{figure}[b]
	\centering
	\includegraphics[width=8.6cm]{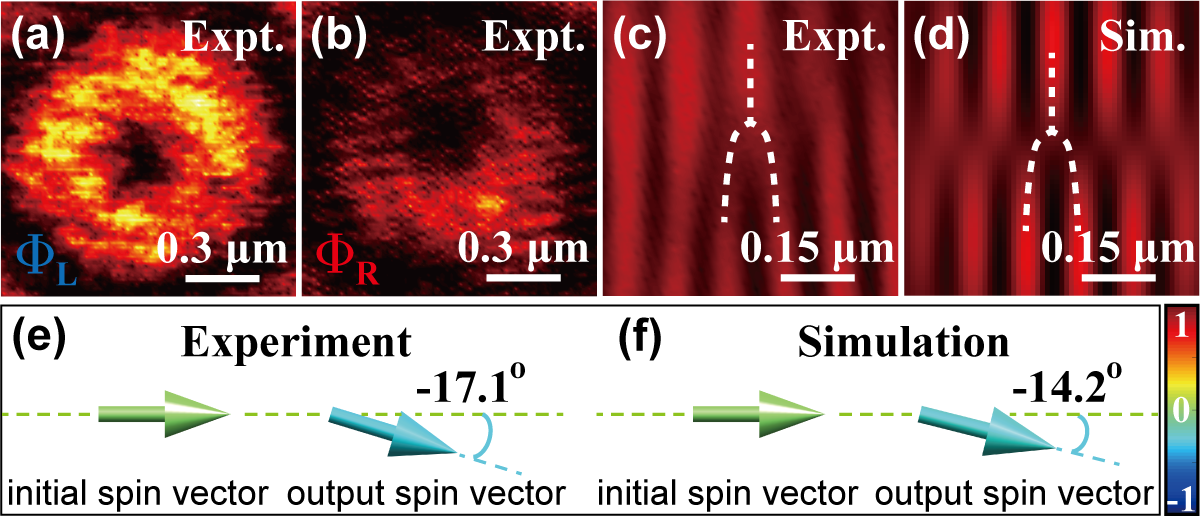}
	\caption{ Observation of the spin rotation by using the deep-subwavelength BG beam ($r_0=0.12$ $\mu$m). (a, b) Experimentally measured intensity distributions of the left- and right-handed circular polarizations. (c, d) Plane-wave interference patterns obtained both in experiment (c) and in simulation (d). (e, f) The measured output spin rotation in comparison with the initial one: (e) experiment; (f) simulation.}
\end{figure}
\indent We observe more prominent spin precession by considering the Bessel structured light with deeper subwavelength feature size. The carrier envelope is replaced by
\begin{equation}
\tilde{A}(r)=J_{\ell}(r/r_0)
\end{equation}
where $J_{\ell}$ denotes the Bessel function of order $\ell$. In practice, we should properly truncate the ideal Bessel beam by using a Gaussian factor. The resultant Bessel-Gaussian (BG) profile exhibits nondiffracting property over a certain distance. We generate this BG beam using a metasurface whose geometry exhibits cylindrical symmetry. The highly localized BG beam is a result of in-phase interference of many high-spatial-frequency waves \cite{Fu2020}. We demonstrate result for a beam parameter of $r_0=0.12$ $\mu$m, while maintaining other parameters unchanged. Similarly, an initial balance between the left and right-handed components is broken by the SOC [Fig. 5(a, b)]. The output spin rotates to a larger angle of -17.1$^{\text{o}}$, nearly in accordance with the theoretical calculation [Fig. 5(f)]. The measured and simulated interference patterns verify the spin-precession-induced OAM generation, see Fig. 5(c) and 5(d), respectively. \\
\begin{figure}[t]
	\centering
	\includegraphics[width=8.6cm]{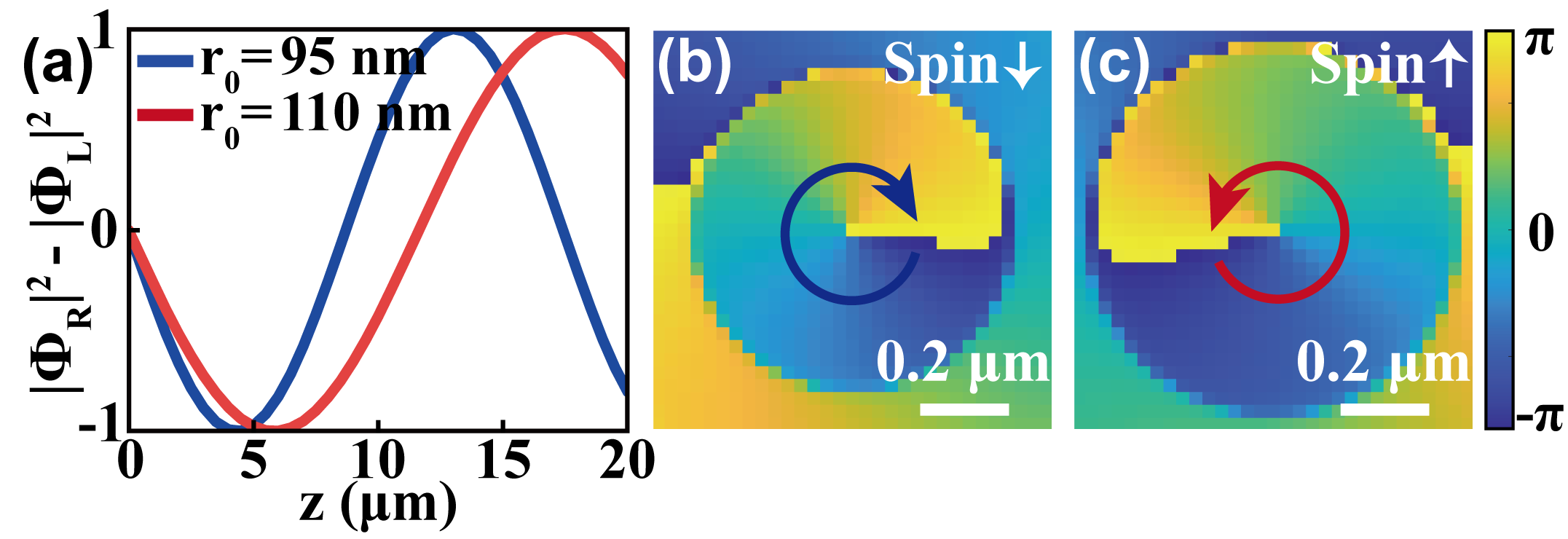}
	\caption{(a) The simulated [based on Eq. (6)] beam-dependent spin oscillatory modes. (b)(c) The simulated [based on Eq. (1)] phase distributions of the output light states from the barium metaborate crystal film ($\bar{\gamma}=-0.16$), for (b) $r_0=95$ nm, and (c) $r_0=110$ nm.}
\end{figure}
\indent Finally, we propose to using the beam-dependent SOC in precision measurement of slight variation of structured light, with measurement accuracy up to 15 nm. This nanometric resolution is usually impossible to be reached by current optical detectors. It requires to realize rapid oscillation between the spin up and spin down. Specifically, we exploit the deep-subwavelength BG beam as carrier envelope of the spin. In this scenario, the Pauli equation [Eq. (6)] emulates a SOC process for the spin oscillation. Figure 6(a) depicts the SOC-supported spin harmonic oscillations along with the coupling distance, for different cases of beam widths. Obviously, the spin oscillation is very sensitive to the change of spatial structure of light, giving rising to ultrasensitive beam-dependent oscillatory modes. As a result, a slight change of the beam width leads to significant spin flipping. This allows to detect the spatial variation of light as small as 15 nm. To verify the result, we present simulated outcomes [see Fig. 6(b) and 6(c)], clearly showing opposite helical wavefronts of the output states (corresponding to the spin down and spin up), for $r_0$=95 nm and $r_0$=110 nm. Note that one can further increase the measuring sensitivity by properly reducing the beam width.
\section{conclusion}
\indent In summary, we have demonstrated both theoretically and experimentally novel SOC phenomena, caused by the deep-subwavelength spin-orbit structured light. This beam-dependent SOC contrasts to those being material-dependent \cite{Mann2020,Szczytko2019}. The reported SOC is closely relevant to the spatial gradient of light field, hence it can be significantly enhanced by using the deep-subwavelength carrier envelopes. We have qualitatively characterized this effect, by measuring the spin precessions under different beam parameters. Particularly, based on the deep-subwavelength Bessel beam, a significant spin rotation about -17.1$^{\text{o}}$, accompanied by OAM generation, was achieved within a coupling length of only one wavelength. The influence of the phase mismatch on the beam-dependent SOC was also discussed, see Appendix D. These fundamental SOC phenomena may find interesting applications in different areas \cite{Mirhosseini2013,JChen2021,Vijay2012,Abele2018}. As an example, we have proposed to use such a strong SOC effect in the precise measurement of slight spatial change of light with nanometric resolution.
\section{ACKNOWLEDGMENTS}
\indent We thank Boris Malomed from Tel Aviv University for kind discussions about the SOC. This work was supported by the National Natural Science Fundation of China (62175091, 12304358), and the Guangzhou science and technology project (202201020061).
\section{APPENDIX A: Analogy of spin-orbit coupling in spin-1/2 system and synthetic two-level system}
\begin{table*}[t]
	\renewcommand\arraystretch{1.5}
	\centering
	\caption{Analogies between the presented synthetic spin-1/2 system in the higher-order optical regime and the spin-1/2 system in the quantum mechanics. The direct analogies between these two different settings enable us to emulate intriguing spin transport phenomena in the presence of spin-orbit coupling.}
	\setlength{\tabcolsep}{5mm}{ 
		\begin{tabular}{ccc}
			\hline
			\textbf{Physical parameters} & \textbf{Spin-1/2 system} & \textbf{Synthetic spin-1/2 system}\\
			\hline
			Spins & $\Phi_{\uparrow}$ and $\Phi_\downarrow$ & $\hat{R}$ and $\hat{L}$ \\
			Eigenstates & $\Phi_{+}^{1/2}$ and $\Phi_{-}^{1/2}$ & $\Phi_{+}$ and $\Phi_{-}$ \\
			Field vector & \textbf{B} (real) & \textbf{B}=$(-\bar\gamma/(\bar\beta\tilde{A})\nabla_{xy}^2\tilde{A}, 0, -\Delta\beta)$ \\
			Spin-orbit coupling term & $\mathbf{H_{1/2}}=\sigma\cdot\mathbf{B}$ & $\mathbf{H_{SOC}}=\sigma\cdot\mathbf{B}$ \\
			Space/time coordinates & $(x,y,t)$ & $(x,y,z)$\\
			Mass & $m$ & $M=2\bar\beta\tilde{A}/(2-\bar\gamma)$\\
			Momentum operator & \textbf{P}$_{\perp}^{2}=[-\nabla_\perp^2, 0; 0, -\nabla_\perp^2]$ & \textbf{P}$_{\perp}^{2}=[-\nabla_\perp^2, 0; 0, -\nabla_\perp^2]$ \\
			\hline
	\end{tabular}}
\end{table*}
\indent The spin-1/2 dynamics in the external vector field \textbf{B} can be described by a Hamiltonian term \(\mathbf{H_{1/2}}\mathrm{=}\sigma*\textbf{B}\), where $\sigma$ is the Pauli matrix vector. In a normalized form, it can be expressed as
\begin{equation}
	\mathbf{H_{1/2}}=\frac12{\begin{bmatrix}\cos\theta&\mathrm{sin}\theta\cdot\mathrm{exp}\left(-i\varphi\right)\\\mathrm{sin}\theta\cdot\exp\left(i\varphi\right)&-\mathrm{cos}\theta\end{bmatrix}}
\end{equation}
where $\theta$ and $\varphi$ are two angles that define a normalized (unit) sphere. The vector \(\mathbf{B}\) then possesses around the sphere, with direction determined by $\theta$ and $\varphi$. This Hamiltonian \(\mathbf{H_{1/2}}\) admits two spin eigenstates that point along to \(\mathbf{B}\), written as 
\begin{equation}
	\begin{gathered}
		\Phi_{+}^{1/2} =\cos\left(\frac\theta2\right)\Phi_\uparrow+\exp(i\varphi)\sin\left(\frac\theta2\right)\Phi_\downarrow  \\
		\Phi_{-}^{1/2} =\sin\left(\frac\theta2\right)\Phi_\uparrow-\exp(i\varphi)\cos\left(\frac\theta2\right)\Phi_\downarrow 
    \end{gathered}
\end{equation}
where \(\Phi_{\uparrow}=[1\quad0]^{\mathrm{T}}\) and \(\Phi_\downarrow=[0\quad-1]^\mathrm{T}\) are spin-up and spin-down states defined in the \(z\) direction. Figure 1(b) geometrically depicts this picture onto a Bloch sphere. All possible spins of the system can now be mapped onto the sphere, with the spin up \(\Phi_{\uparrow}\) and spin down \(\Phi_\downarrow\) located at the north and south poles of the sphere, respectively. In the presence of the external field \(\mathbf{B}\), the initial spin precesses around the vector \(\mathbf{B}\), giving rise to many intriguing spin transport phenomena such as the geometric phase.\\  
\indent In our case, we study spin-orbit coupling of structured light in a photonic crystal. The structured light in the system is comprising a superposition of two orthogonal spin-orbit states with non-trivial topological structures. They can be written as \(\hat{R}=\exp{(il\phi)}(\hat{x}-i\hat{y})/\sqrt{2}\) and \(\hat{L}=\exp{(-il\phi)}(\hat{x}+i\hat{y})/\sqrt{2}\), respectively. These topological states define the spin up and spin down equivalents along the $z$ axis, respectively, but they are not eigenstates of the analogous spin-orbit Hamiltonian \(\mathbf{H_{soc}}=-\sigma*\mathbf{B}\). In the circular basis, a similar Hamiltonian matrix can be written as 
\begin{equation}
	\mathbf{{H}_{soc}}=\begin{bmatrix}B_2&B_3-iB_1\\B_3+iB_1&-B_2\end{bmatrix}
\end{equation}
In our case, since $B_2$ is zero (see the main text), the effective vector \textbf{B} obtained here lies on the purely transverse $B_1$$B_3$ plane, as shown in Fig. 1(a). As a result, the pseudospin eigenstates of \(\mathbf{{H}_{soc}}\) that point along this transverse vector  \textbf{B}  comprise an equal superposition of $\hat{R}$ and $\hat{L}$, express as  
\begin{equation}
	\begin{aligned}\Phi_+&=\cos\left(\frac\pi4\right)\hat{R}+\exp(i\varphi)\sin\left(\frac\pi4\right)\hat{L}\\\Phi_-&=\sin\left(\frac\pi4\right)\hat{R}-\exp(i\varphi)\cos\left(\frac\pi4\right)\hat{L}
	\end{aligned}
\end{equation}
We can now interpret these eigenstates as a mixing of $\hat{R}$ and $\hat{L}$. \text{Poincaré}-sphere representation allows us to visualize these spin eigenstates as well as the pure states $\hat{R}$ and $\hat{L}$. Clearly, this is analogous to the Bloch-sphere representation for the spin-1/2 system. The spin-orbit coupling makes this state evolves along the \text{Poincaré} sphere, which can be described by the synthetic Pauli equation, 
\begin{equation}
	i\frac{\partial\Phi}{\partial z}=\left(\frac1{2M}\mathbf{P}_\perp^2\tilde{A}-\frac12\sigma\cdot\mathbf{B}\right)\Phi
\end{equation}
Table I summaries analogous formulas between these two systems.
\section{APPENDIX B: Theoretical derivation for the topology-preserving flat lens}
\indent In this section, we theoretically prove that the flat lens used in the experiment does not change the spin-orbit property of the LG beam after tightly focusing. The flat lens is designed by an amplitude-only hologram generated from an interference between an angular cosine wave and a spherical wave (see ref. [61] in the text). When the LG beam \(\tilde{A}(x,y)\) carrying a general spin state $\Phi$ passes through the flat lens, it is modulated in binary. As a result, the light field behind the flat lens can be expressed as 
\begin{equation}
	\mathbf{E}(x,y,z=0)=\tilde{A}(x,y)*t(x,y)\big[\Phi_x(\phi)\hat{x}+\Phi_y(\phi)\hat{y}\big]
\end{equation}
where \(t(x,y)\) denotes transmission function of the flat lens and \(\phi=\arctan\left(y/x\right)\). Within this initial condition, we solve the diffractive problem according to the vectorial Helmholtz wave equation. The diffractive field at the focal plane of the flat lens can be written as

\begin{equation}
	\begin{split}
	\mathbf{E}(x,y,z_{f})  = & \frac{k}{i2\pi z_{f}}\iint\mathbf{E}(x^{\prime},y^{\prime},z=0)
	\\ & \exp\left\{\frac{ik}{2z_{f}}[(x-x^{\prime})^2+(y-y^{\prime})^2]\right\}dx^{\prime}dy^{\prime}
\end{split}
\end{equation}
Note that owing to the cylindrical symmetry of the flat lens (see the layout in the text, Fig. 2(b)), the transmission function can be also given in a cylindrical form of \(t(r)\), where \(r=(x^{2}+y^{2})^{1/2}\). In this case, the complex amplitude of the initial field is separable in the polar coordinates \((r,\phi)\). We therefore rewrite the solution in the cylindrical coordinate system and deal with the integrals. We finally obtain the analytical solution for the vectorial light field at the focal plane, given by 
\begin{equation}
		\mathbf{E}(x,y,z_{f}) = f(r)\big[\Phi_x(\phi)\hat{x}+\Phi_y(\phi)\hat{y}\big]
\end{equation}
where
\begin{equation}
	\begin{split}
	f(r) = & -\frac k{z_f}\int_0^\infty\tilde{A}(r^{\prime})t(r^{\prime})r^{\prime}J_1\left(\frac{krr^{\prime}}{z_f}\right)
	\\ & \exp\left[\frac{ik}{2z_f}(r^2+r'^{2})\right]\mathrm{d}r^{\prime}
\end{split}
\end{equation}
and \(J_1\) indicates the first-order Bessel function. It is evident that the diffractive field at the focal plane shares a similar analytic form to the initial one, except for that the envelope becomes a \(z\)-dependent function. It indicates that the flat lens can completely retain the initial spin state when it is focused into the input end of the crystal. The topology-preserving flat lens enables us to detect the pseudo spin precession caused by the deep-subwavelength structured light, which cannot be achieved by using the conventional high NA objective lens. 
\section{APPENDIX C: Purity measurement of the first-order LG beam from the q-plate}
\begin{figure}[t]
	\centering
	\includegraphics[width=8.6cm]{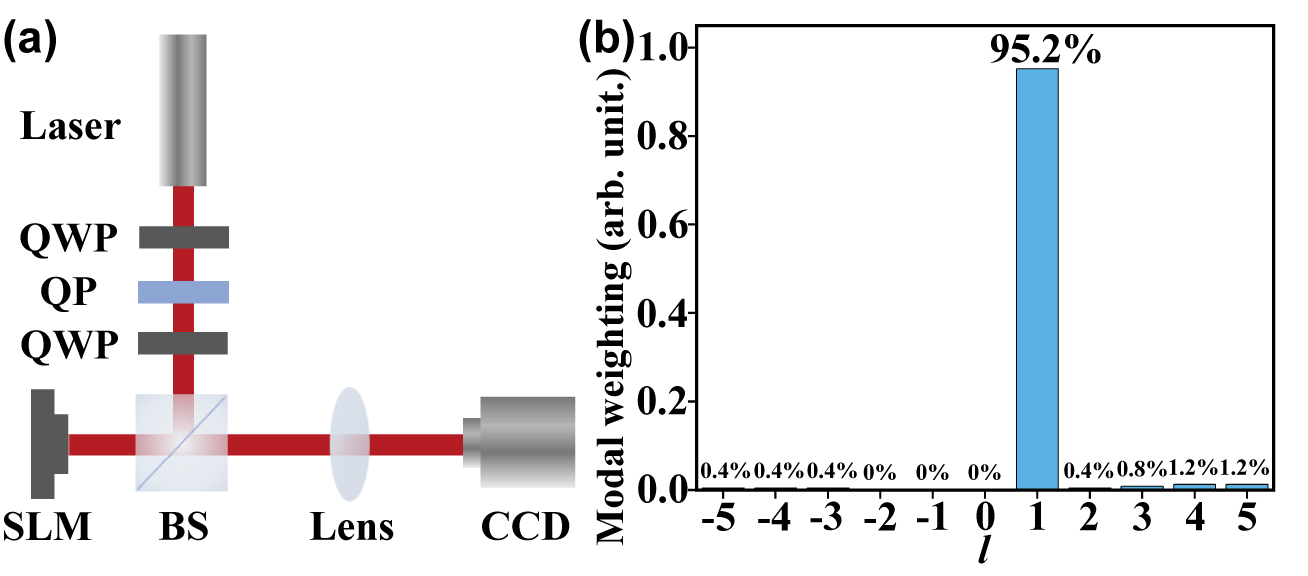}
	\caption{Modal decomposition results. (a) An experimental setup used to measure the purity of the first-order LG beam emerging from the q-plate. The LG beam is decomposed into LG basis modes. The linearly polarized He-Ne laser operating at the wavelength of 632.8 nm is considered. QWP, quarter wave plate; QP, q-plate with a topological number of \(q=1/2\); SLM, spatial light modulator; BS, beam splitter; CCD, charge-coupled device. (b) The modal decomposition results at the basis of LG modes with topological charge ranging from \(l=\) -5 to 5.}
\end{figure} 
\indent we perform additional experiment to show that the generated first-order LG beam from the q-plate is of high purity, which is sufficiently enough to detect the photonic spin-orbit coupling effect. We utilize a modal decomposition method \cite{schulze2013,wei2019} to measure the purities of the output LG mode from the q-plate with a topological charge of \(q=1/2\), see an experimental setup in Fig. 7(a). Two quarter wave plates (QWPs) are used to select a proper polarization of the generated first-order (\(l\)=1) LG beam that matches to the spatial light modulator (SLM). A group of pure LG modes generated from digital holograms by using the SLM are considered to decomposed the LG beam. Fig. 7(b) shows the decomposing result depicted in a histogram. It is seen that the measured purity of the first-order LG beam from the q-plate is 95.2$\%$.
\begin{figure}[t]
	\centering
	\includegraphics[width=8.6cm]{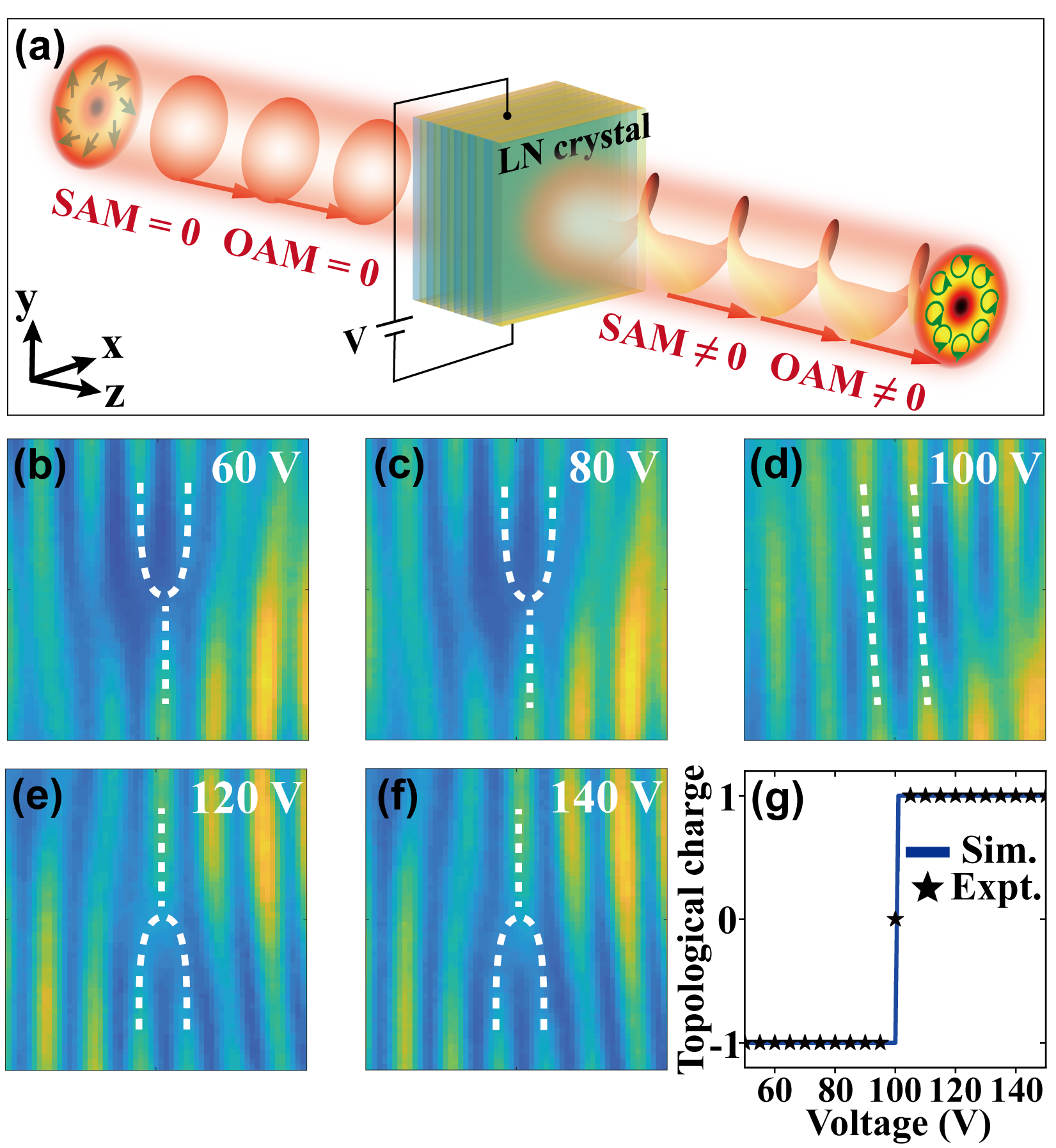}
	\caption{Controllable spin-orbit coupling by engineering the phase mismatch in a c-cut electro-optic lithium niobate crystal. (a) Experimental scheme for observing the electrically engineered spin-orbit coupling. (b-f) Experimentally measured photonic spin states at different applied voltages: (b) $U$=60 V, (c) $U$=80 V, (d) $U$=100 V, (e) $U$=120 V, and (f) $U$=140 V. (g) The measured topological charge as a function of applied voltage. Sim: simulation; Expt: experiment. In this experiment, the coupling length of the crystal is set to $z$=30 $mm$. }
\end{figure}
\section{APPENDIX D: Engineering photonic spin-orbit coupling by tuning the phase mismatch}
\indent In addition to the beam-dependent photonic spin-orbit coupling which we have shown in the main text, we perform additional experiments confirming that the spin-orbit coupling can be also controlled by engineering the phase mismatch. To this end, we consider electrically tuning the phase mismatch in a c-cut electro-optic lithium niobate (LN) crystal, whose optical axis is in accordance with propagation direction of the beam [see Fig. 8(a)]. In the presence of transverse modulation, the phase mismatch can be written as
\begin{equation}
	\Delta\beta=-k_0n_o^3\gamma_{22}U/d
\end{equation}
where \(k_{0}=2\pi/\lambda \) denotes wavenumber in vacuum with \(\lambda\) being wavelength, \(n_o\) is the ambient refractive index of the crystal, $U$ is the applied voltage, \(d\) is the thickness, and \(\gamma_{22}=6.8\) \(\mathrm{pm/V}\) is an electro-optic coefficient of the crystal. In this case, the external knob \(U\) is utilized to finely tune the phase mismatch and the resulting spin-orbit coupling. We use the same experimental setup and obtain a voltage-dependent transition between different spin states in the phase mismatching regime. Panels (b-f) in Fig. 8 show controllable spin states of light by varying the applied voltage. Moreover, we perform detailed experiments to measure the topological charge of the output light state as a function of voltage, see Fig. 8(g). These results suggest another important degree of freedom for engineering the spin-orbit coupling.


\begin{thebibliography}{99}
\bibitem{SR} In experiment, the spin rotation was obtained by measuring $S_2=|\Phi_R|^2-|\Phi_L|^2$. Ideally, the spatial distribution of $S_2$ is homogeneous; whereas in experiment it exhibits fluctuations due to the imperfect beam envelope. We therefore average it by using an expression: $\bar{S}_2=\sum_n^{N_x}\sum_m^{N_y}S_{2,n,m}/(N_xN_y)$.
\end{thebibliography}
\end{document}